\documentclass[
	aps,
	showpacs,
	showkeys,
	twocolumn,
	altaffilletter,
	nolongbibliography,
	numerical,
	flushbottom,
	secnumarabic,
	pra,
	superscriptaddress,
	floatfix,
	10pt
]{revtex4-1}

\usepackage{lipsum}

\usepackage[dvipsnames]{xcolor}

\usepackage{bm}

\usepackage{graphicx}

\usepackage{upgreek}

\usepackage{natbib}

\usepackage{braket}

\usepackage{physics}

\usepackage{newtxtext, newtxmath}

\usepackage{siunitx}

\usepackage[caption=false]{subfig}
\usepackage[%
	breaklinks,%
	pdftex,%
	hyperfootnotes=true,%
	pdfpagelabels,%
	bookmarks,%
	pageanchor,%
]{hyperref}

\hypersetup{%
	colorlinks=true, linktocpage=true, pdfstartpage=1, pdfstartview=FitH, pdfborder={0 0 0},%
	breaklinks=true, pdfpagemode=UseNone, pageanchor=true, pdfpagemode=UseOutlines,%
	plainpages=false, bookmarksnumbered, bookmarksopen=true, bookmarksopenlevel=1,%
	hypertexnames=true, pdfhighlight=/O,
	urlcolor=Red!50!Sepia, linkcolor=NavyBlue, citecolor=RoyalBlue, 
}

\graphicspath{{Pictures/}}

\newcommand{\ie}{i.\,e.}

\newcommand{\eg}{e.\,g.}
\newcommand{\ped}[1]{_\text{#1}}
\newcommand{\api}[1]{^\text{#1}}

\newcommand{\ham}{H}

\newcommand{\rng}[2]{\ensuremath{[#1, #2]}}
\renewcommand{\epsilon}{\varepsilon}

\newcommand{\beq}{\begin{equation}}
\newcommand{\eneq}{\end{equation}}

\DeclareMathOperator{\eu}{e}
\DeclareMathOperator{\iu}{i}

\begin{document}

\author{L.~M.~Cangemi}
\affiliation{Dipartimento di Fisica ``E.~Pancini'', Universit\`a di Napoli ``Federico II'', Complesso di Monte S.~Angelo, via Cinthia, 80126 Napoli, Italy}
\email{lorismaria.cangemi@unina.it}
\affiliation{CNR-SPIN, c/o Complesso di Monte S. Angelo, via Cinthia - 80126 - Napoli, Italy}
\author{V.~Cataudella}
\affiliation{Dipartimento di Fisica ``E.~Pancini'', Universit\`a di Napoli ``Federico II'', Complesso di Monte S.~Angelo, via Cinthia, 80126 Napoli, Italy}
\affiliation{CNR-SPIN, c/o Complesso di Monte S. Angelo, via Cinthia - 80126 - Napoli, Italy}
\author{M.~Sassetti}
\affiliation{Dipartimento di Fisica, Universit\`a di Genova, Via Dodecaneso 33, 16146 Genova, Italy} 
\affiliation{CNR-SPIN,  Via  Dodecaneso  33,  16146  Genova, Italy}
\author{G.~De Filippis}
\affiliation{Dipartimento di Fisica ``E.~Pancini'', Universit\`a di Napoli ``Federico II'', Complesso di Monte S.~Angelo, via Cinthia, 80126 Napoli, Italy}
\affiliation{CNR-SPIN, c/o Complesso di Monte S. Angelo, via Cinthia - 80126 - Napoli, Italy}

\title{Dissipative dynamics of a driven qubit: interplay between non-adiabatic dynamics and noise effects from weak to strong coupling regime}

\date{\today}


\begin{abstract}
We study the exact solution of the Schr\"odinger equation for the dissipative dynamics of a qubit, achieved by means of Short Iterative Lanczos method (SIL), which allows us to describe the qubit and the bath dynamics from weak to strong coupling regimes. 
We focus on two different models of a qubit in contact with the external environment: the first is the Spin Boson Model (SBM), which gives a description of the qubit in terms of static tunnelling energy and a bias field. The second model describes an externally driven qubit, where both the bias field and the tunnelling rate are controlled by a time-dependent magnetic field obeying to a finite time protocol. We show that in the SBM case, our solution correctly describes the crossover from coherent to incoherent behavior of the magnetization, occurring at the Toulouse point. Furthermore, we show that the bath response dramatically changes during the system dynamics, going from non-resonant at small times to resonant behavior at long times. When the external driving field is present, for fixed values of the drive duration our results show that the bath can provide beneficial effects to the success of the protocol.  
We find evidence for a complex interplay between non-adiabaticity of the protocol due to the external drive and dissipation effects, which strongly depends on the detailed form of the qubit-bath interaction.   
       
\end{abstract}

\maketitle

\section{Introduction}\label{sec:intro}
     
     Models of quantum systems interacting with their environment are of primary importance in the field of open quantum systems \cite{breuer:open-quantum,Weiss:open-quantum2}. In the last decades, several experimental achievements, both in the field of quantum information and quantum simulation, have stimulated a renewed interest in these problems \cite{harris:d-wave}. 
     
     In Adiabatic Quantum Computation (AQC) protocols, the interaction of a system of mutually coupled qubits with a thermal bath can lead to noticeable changes in the mechanism of defects creation, affecting the success probability of the computation \cite{Smelyanskiy:decoherence,Keck:dissipationadiab}.~However, decoherence effects are not necessarily detrimental to the success of AQC \cite{childs:robustness,albash:decoherence}; further, it has also been shown that the interaction with an external bath at finite temperature can have beneficial effects on the success of AQC \cite{amin:thermal-qa,dickson:thermal-qa,Gianluca}, though the underlying mechanism providing such an advantage over closed-system dynamics, as well as its dependence on the nature of quantum critical point 
     remains rather unclear \cite{arceci:dissipative-lz,mishra:thermal-gap-qa}. 
     
     Several prototypical models of driven open quantum systems \cite{Grifoni:drivenquantum} can prove useful in the study of out-of-equilibrium quantum thermodynamics \cite{SassettiWeiss:energyexchange, Esposito:Noneq1,Campisi:Noneq2}, a well-estabilished field that in the last decades has regained popularity since the discovery of its links with quantum information theory \cite{Goold:entropyandinfo}. In this context, entropy production in systems interacting with a heat bath as well as with finite, engineered baths has been the focus of recent experimental and theoretical work \cite{Brunelli:entropyexp,Manzano:entropy}. Moreover, the study of energy and heat exchanges mechanism between an externally driven quantum system and one or more heat baths is relevant for the theoretical understanding of quantum heat engines \cite{CampisiPekolaFazio:quantheng,Manzano:entropy2}.  
     
     The prototypical model of a two level system (TLS) interacting with a thermal bath, also known as Spin Boson Model (SBM) has been considered in several works, aimed at describing its out-of-equilbrium dynamics and quantum phase transition arising from different kinds of dissipation \cite{caldeira:caldeira-leggett,Leggett,Sassetti:univ,LeHurNRG1,LeHurImpurity}. Energy exchange in dissipative driven TLS has also been considered \cite{SassettiWeiss:energyexchange} from weak to strong coupling regimes. Further, the effect of the environment on the ground state topology of SBM has been studied \cite{henriet:Topologysbm}, showing that only local geometric properties are noticeably affected, while global properties remain unchanged as long as the system is in the delocalized phase, \ie~the coupling to the bath degrees of freedom does not exceed the critical value ($\alpha < \alpha\ped{c}=1$). 
     
     In addition, the bath-induced non-adiabaticity has been addressed \cite{henriet:Topologysbm}, and at strong coupling regime the crossover from quasi-adiabatic to non-adiabatic dynamics due to the environment has been studied. While this picture holds true in the quasi-adiabatic regime, it is not a priori clear how the environment affects the dynamics of the TLS at not-so-low sweep velocities; furthermore, different forms of the coupling could lead to changes in this scenario, as stressed in several works \cite{thorwart1,thorwart4,arceci:dissipative-lz} addressing the dissipative dynamics of Landau-Majorana-St\"uckelberg-Zener (LMSZ) model \cite{landau:crossings,Majorana1932,zener:crossings}. 
     
     In this paper, we study the dissipative dynamics of a two level system, \ie~a qubit subject to external driving fields and interacting with its environment, from weak to strong coupling regimes. We first address the static field case, 
     focusing on the biased SBM in contact with an Ohmic bath, a widely-studied model which describes the effects of dissipation and decoherence on a TLS. In addition, we consider a time-dependent protocol, which has been recently implemented in solid-state devices in order to realize dynamical measurements of topological phase transitions \cite{Gritsev,RoushanNature:topotrans}. 
     
     We employ an exact numerical approach based on a truncation scheme of the bath Hilbert space and on the Short Iterative Lanczos (SIL) diagonalization \cite{DeFilippisOc,DeFilippisSpinLattice,nature:Giulio,letters:DeFilippisPumpProbe,letters:DeFilippisSharpTrans,Lanczos1,Lanczos2}, which allows us to follow the dynamics of the observables of both the qubit system and the environment, without the need of tracing out the bath degrees of freedom. This method has proved useful in reproducing the correct physical behavior of the SBM in the weak coupling regime \cite{Cangemi:SIL}, and we show how the inclusion of higher order excitation processes in the physical description can noticeably widen the range of coupling strengths to be investigated, allowing us to describe the physics from intermediate to strong coupling regime where no analytical scheme is known to hold. 
     
     We show that in the case of unbiased SBM our approach is successful in describing the crossover from coherent to incoherent behavior of magnetization dynamics, occurring at the Toulouse point at $\alpha=1/2$. In addition, taking advantage of our technique we find the dynamical evolution of the mean population of the bath modes as a function of time, and we observe a change from non-resonant to resonant response at fixed coupling strengths. Furthermore, in the case of the driven qubit, we show a non-monotonic behavior of the fidelity at fixed final times as a function of the dissipation strength: this behavior, which is found to depend on the detailed form of interaction with the environment, signals the complex interplay between non-adiabatic effects due to the external time-dependent driving force and dissipation.      
     
     The paper is organized as follows: in section \ref{sec:modelHam} we introduce the general Hamiltonian scheme we intend to study, focusing on the characteristic form of coupling with the bath. In section \ref{sec:Toulouse},\ref{sec:Magnetic} we discuss two different prototypical models of a TLS interacting with external environment, subject to static and time-dependent external fields, and we analyze numerically their open dynamics from weak to strong coupling regimes. We test our predictions by a comparison with well known theoretical approximations, and discuss their possible physical intepretations. Eventually, in \ref{sec:conclusions} we discuss viable extensions of this work along with future perspectives.             
                            
     \section{Model Hamiltonian}\label{sec:modelHam}
    
     We focus on a TLS system, \ie~a qubit subject to time-dependent external fields, which is interacting with its environment. The qubit is described by a time-dependent Hamiltonian $H\ped{S}(t)$. The total Hamiltonian of the interacting system is written as
     \begin{equation}\label{eq:TotalHam}
       H(t)= H\ped{S}(t) + H\ped{B} + H\ped{I}, 
     \end{equation}  
     where $H\ped{B}$ is the free Hamiltonian of the bath and $H\ped{I}$ is the coupling energy between the qubit and the bath. Making use of the spin $1/2$ Pauli matrices $\boldsymbol{\sigma}=(\sigma_x,\sigma_y,\sigma_z)$, the qubit Hamiltonian can be written in a very simple form,  
     \begin{equation}\label{eq:QubitHam}
     H\ped{S}(t)=-\frac{1}{2}\boldsymbol{h}(t)\cdot \boldsymbol{\sigma},  
     \end{equation}      
     where we take as $\boldsymbol{h}$ a time-dependent magnetic field vector which at fixed time $t$ points in a given direction of the three-dimensional coordinate space. We conventionally adopt as a basis for the qubit states, \ie~the computational basis, the set of eigenstates of $\sigma_z$ operator, namely $\sigma_z\ket{\hat{z}; \pm}=\pm\ket{\hat{z}; \pm}$: as a consequence, the  component of $\boldsymbol{h}$ along the $\hat{z}$ axis acts as a bias on the energy levels of the two states, while linear combinations of $\sigma_\pm$ operators give rise to tunnelling between these two states. 
     
     As usual, we model our bath by means of a collection of bosonic oscillators of frequency $\omega_k$, and the Hamiltonian $H\ped{B}$ can be conveniently written in terms of creation (annihilation) operators $b^{\dagger}_k (b^{\phantom{\dagger}}_k) $ obeying to bosonic commutation relations $\comm{b^{\dagger}_k}{b^{\phantom{\dagger}}_l}=\delta_{kl}$, 
     \begin{equation}\label{eq:BathHam}                
     H\ped{B}=\sum_k\omega_k b^{\dagger}_k b_k, 
     \end{equation}
     The time-independent interaction term couples the qubit with the external bath along a given direction $\hat{n}$ as follows
     \begin{equation}\label{eq:IntHam}
     H\ped{I}=\frac{1}{2}\boldsymbol{\sigma}\cdot\boldsymbol{\hat{n}}\sum_k \lambda_k (b^{\dagger}_k + b_k ), 
     \end{equation}    
     where $\lambda_k$ is the coupling strength with the k-th oscillator. The bath properties are described by its spectral density $J(\omega)$ \cite{Leggett}: it can be written as a sum over discrete frequencies of the bath modes, ranging from $0$ up to a cutoff frequency $\omega\ped{c}$, which is the greatest energy scale of the system. In the continuum limit, the spectral density takes the form         
     \begin{equation}\label{eq:SpectDensity}
     J(\omega)= \sum_k \lambda_k^2 \delta(\omega-\omega_k)= 2 \alpha \frac{\omega^s}{\omega^{s-1}_c}\eu^{-\frac{\omega}{\omega\ped{c}}},
     \end{equation}  
     Here the adimensional parameter $\alpha$ measures the strength of the dissipation, while the parameter $s$ distinguishes among three different kinds of dissipation that have been studied in the recent literature \cite{Weiss:open-quantum2,Leggett,bulla:nrg2}: Ohmic ($s=1$), sub-Ohmic ($s<1$) and super-Ohmic case ($s>1$). 
     The expression for the coupling term in Eq.~\eqref{eq:IntHam} is rather general: it has been proposed to study the effect of a thermal environment on qubits subject to different time-dependent protocols, including the widely studied LMSZ sweeps \cite{Vavilov:Decoherence,thorwart4,arceci:dissipative-lz}; in the latter case, it has been argued that the introduction of a "transverse" coupling direction, \ie~orthogonal to the time-dependent bias field, could provide a simple theoretical explanation of the experimental findings regarding D-Wave Rainier's chip \cite{dickson:thermal-qa}. 
     
     In the following we extend our analysis to a qubit coupled to a thermal bath along different spatial directions. As a first example, we consider the dissipative dynamics of a widely studied model in the spin-boson literature, which accounts for decoherence and dissipation effects on the qubit dynamics \cite{Leggett,LeHurImpurity,LeHurNRG1}; here the qubit is coupled with the heat bath along $\hat{z}$ axis, playing the role of an additional bias field on the computational basis states $\ket{\hat{z}; \pm}$; if the external fields are time-independent, in the case of Ohmic dissipation ($s=1$), as the coupling strength reaches its critical value $\alpha_c=1$ this model predicts the occurrence of a Quantum Phase Transition (QPT) of Kosterlitz-Thouless kind. As a second example, we analyze a time-dependent protocol where the qubit is subject to a rotating magnetic field $\boldsymbol{h}(t)$ performing a sweep in a fixed plane, and the dissipation can take place along two particular directions in the plane of rotation.

	\section{Spin Boson model}\label{sec:Toulouse}
	
	We study the dynamics of model in Eq.~\eqref{eq:TotalHam}, taking a static tunnelling element along $\hat{x}$ axis, \ie~$h_x=\Delta$, a bias field along $\hat{z}$ axis, $h_z=h_0\mbox{, } h_y=0$, and restricting to the case of Ohmic dissipation ($s=1$), with $\boldsymbol{\hat{n}}=\hat{z}$ : hence our model reduces to the biased SBM. 
	
	In the last decades, several works have been written in order to characterize its quantum phase diagram and describe the corresponding dynamical properties under different parameter regimes. A number of approximate analytical treatments have been devised in order to compute the behavior of one of the most important correlators as a function of time, \ie~the magnetization along the $\hat{z}$ axis $\ev{\sigma_z(t)}$, which is directly related to experiments. 

	In the unbiased case ($h_0=0$), theoretical approaches based on Conformal Field Theory (CFT) predict that $\ev{\sigma_z(t)}$ exhibits underdamped oscillations in time, for fixed values of the coupling strength $0\leq\alpha<1/2$; however, the detailed expressions of the oscillation frequency and the damping rate depend on the adopted approximation scheme \cite{orth-lehur}. One of the most popular approximate treatments is the Non-Interacting Blip Approximation (NIBA) \cite{Leggett,Grifoni:drivenquantum}: according to NIBA scheme, at $T=0$ the underdamped oscillation in time can be described in terms of analytical functions of the inverse time scale $\Delta\ped{eff}^{-1}$, with the effective tunnelling energy $\Delta\ped{eff}$ reading 
	\begin{equation}
	 \Delta\ped{eff}=[\Gamma(1-2\alpha) \cos\pi\alpha]^{\frac{1}{2(1-\alpha)} }\Delta\ped{r}\mbox{ , } \Delta\ped{r} = \Delta \left(\frac{\Delta}{\omega\ped{c}}\right)^{\frac{\alpha}{1-\alpha}}	
	\end{equation}
	 where $\Delta\ped{r}$ is the renormalized gap \cite{Leggett,orth-lehur}. While the resulting expression for the oscillation frequency is expected not to be valid \cite{EggerCrossover}, this analytical treatment provides the correct result for the quality factor of the damped oscillation (see App.~\ref{app:Toul} for details), which is a monotonic decreasing function of $\alpha$. 
	 
	Although the NIBA approach successfully describes the main features of the system in the unbiased case, in the presence of an external bias field $h_0$, it fails in describing the long-time limit of $\ev{\sigma_z(t)}$. Several exact analytical treatments based on perturbation theory are known which, in the weak coupling limit, give the correct results for the qubit observables, taking into account the fully-quantum correlations between the qubit and bath degrees of freedom. These theories have been employed to derive analytical results for the dynamics of heat exchanges of the qubit with the reservoir \cite{Carrega15:heatexchange}.       
	
	 The SBM problem can be analytically solved for the coupling strength $\alpha = 1/2$ (Toulouse limit). Here a crossover from coherent to incoherent dynamics takes place: at this point, at zero bias, the oscillation frequency tends to vanish, as well as the quality factor. The analytical solution can be found by mapping the SBM into a Resonant Level Model (RLM), describing a single localized impurity in contact with a bath of spinless fermions at the Fermi level; in addition, Coulomb interactions between the impurity and the fermionic bath are present \cite{Leggett,SassettiWeiss:SpinBoson,LeHurNRG1}, and the resulting coupling strength is a function of $\alpha$. For $\alpha = 1/2$, Coulomb interactions vanish and the model can be exactly solved: by preparing the qubit in the state $\ket{\hat{z}; +}$, \ie~$\ev{\sigma_z(0)}=1$, in the limit of small $\Delta/\omega\ped{c}$, \ie~$\omega\ped{c}\to\infty$ the magnetization dynamics takes the form
	
	\begin{equation}\label{eq:Toulouse}
	\ev{\sigma_z(t)}=e^{-\gamma t} + 2 \int_0^t d\tau\frac{\sin(h_0 \tau)}{\beta \sinh(\frac{\pi \tau}{\beta}) }\left( e^{-\gamma \frac{\tau}{2}}-e^{-\gamma t}e^{\gamma \frac{\tau}{2}}\right)
	\end{equation}
	where $\beta=1/T$ ($k\ped{B}=1$) and the damping rate is proportional to the renormalized gap at $\alpha=1/2$, \ie~$\gamma=\pi \Delta^2/2 \omega_c$. In the absence of external bias, $\ev{\sigma_z(t)}$ takes the exponential form which is also recovered in the NIBA approximation.
	
	Different numerical approaches have been devised in order to compute the real time dynamics of this problem in the range of coupling strengths $0 <\alpha < 1/2$, where no exact analytical solution is known to exist \cite{PathInt1,PathInt2,quapi,bulla:nrg1,orth-lehur,Strathearn:SBMMPS}. Here we apply the numerical SIL technique (see App.~\ref{app:SIL}) in order to describe dynamics of the reduced density matrix of the qubit up to the Toulouse point. As stressed in \cite{Cangemi:SIL}, this technique allows us to simulate the exact dynamical evolution of the whole system density matrix $\rho(t)$ in a suitably truncated bath Hilbert space: hence, the observables of the whole qubit + bath system can be computed. 
	We start by preparing the system and the bath at initial time $t\ped{0}$ in a factorized state:
	\begin{equation}\label{eq:startcond}
	\rho(t\ped{0})=\rho\ped{S}(t\ped{0})\otimes \frac{e^{-\beta H\ped{B}}}{Z\ped{B}}
	\end{equation}       
	where $\rho\ped{S}(t)$ is the reduced density matrix of the qubit that can be computed by tracing out the bath degrees of freedom, \ie~$\rho\ped{S}(t)=\tr\ped{B}{\rho(t)}$. We choose $\ket{\hat{z}; +}$ as the initial state of the qubit, while the bath state is taken as the equilibrium state at $T=0$. We model our bath with a collection of $M=50$ bosonic modes, choosing the absolute maximum number of excitations up to $N\ped{ph}=6$ (see App.~\ref{app:SIL} for details) and we fix the cutoff frequency of the bath to $\omega\ped{c}=5 \Delta$. In the following, we restrict to the unbiased case ($h_0=0$), while in App.~\ref{app:Biased}, \ref{app:Toul}, we discuss the biased case along with a comparison with analytical results.
	
\begin{figure}[tb]
		\centering
		\includegraphics[width=\linewidth]{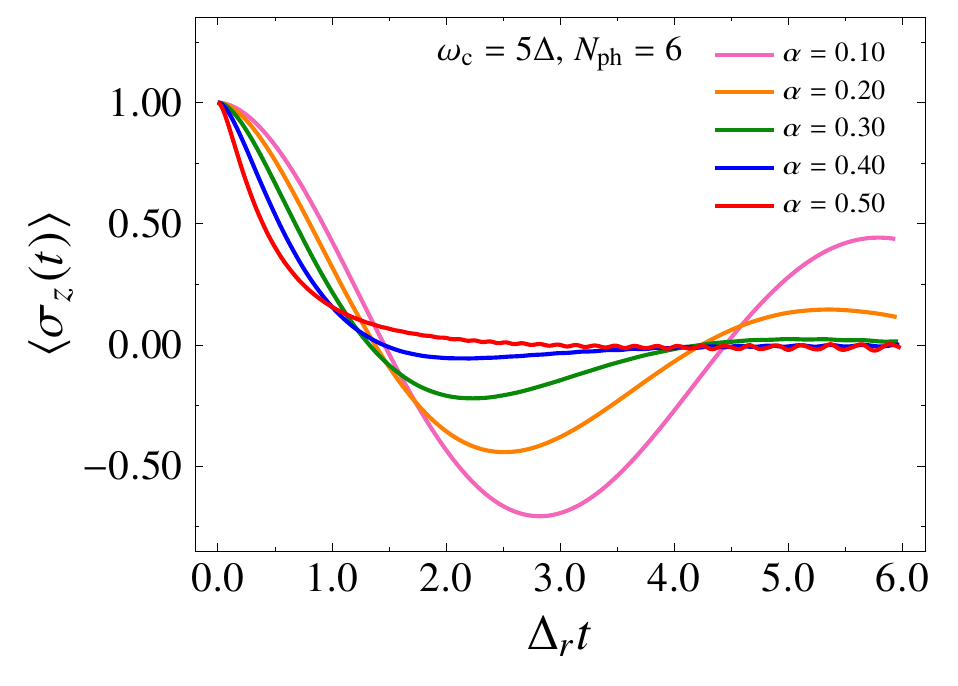}
		\caption{Magnetization $\ev{\sigma_z(t)}$ as a function of the rescaled time $\Delta\ped{r}t$, in the case of Ohmic bath ($ s = 1 $), $T=0$, $h_0=0$, for different values of the coupling strengths $\alpha$ in the range $\numrange[range-phrase = \text{~to~},exponent-product=\cdot]{1e-1}{5e-1}$. The number of bath modes is $M=50$, the cutoff frequency $\omega\ped{c}=5\Delta$ and the maximum number of excitations is $N\ped{ph}=6$.}
		\label{fig:sigmazt}
	\end{figure}

	We simulate the dynamics of the system for different values of the coupling strength $\alpha$, ranging from $\numrange[range-phrase = \text{~to~},exponent-product=\cdot]{1e-1}{5e-1}$. In Fig.~\ref{fig:sigmazt}, we plot the qubit magnetization $\ev{\sigma_z(t)}$ as a function of the rescaled time $\Delta\ped{r} t$: we show that the magnetization dynamics experiences a crossover from a regime of underdamped oscillations in time to an incoherent regime where the oscillation frequency tends to vanish, as long as the coupling strength approaches the expected crossover value $\alpha=0.50$.     
    The crossover from coherent to incoherent behavior can be interpreted in terms of the growth of the entanglement between the qubit and its bath \cite{LeHurNRG1}, a mechanism which can be found in several bipartite systems \cite{AmicoFazioFalci:Entanglement}. 
   
	Starting from the initial condition in Eq.~\eqref{eq:startcond}, where the state of the system is factorized into a product of states of the two subsystem, the state of the qubit thermalizes towards the equilibrium state of the whole Hamiltonian in Eq.~\eqref{eq:TotalHam} at $T=0$, showing entanglement with the bath degrees of freedom. 
	Each numerical curve reported in Fig.~\ref{fig:sigmazt} describes the correct dissipative behavior of the qubit, as it can be shown by a direct comparison with the theoretical result for the quality factor as a function of the coupling strength $\alpha$ (see App.~\ref{app:Toul}). Our results are also in good agreement with recent findings obtained through novel numerical approaches based on non-perturbative techniques \cite{orth-lehur}.
	A detailed comparison with the case $\alpha=1/2$ in Eq.~\eqref{eq:Toulouse}, reported in App.~\ref{app:Toul}, shows also good agreement with theory at long times, while at shorter times small deviations start to appear: it can be explained by the small cutoff value chosen $\omega\ped{c}=5\Delta$, which cannot meet the parameters conditions ensuring the validity of Eq.~\eqref{eq:Toulouse}; we argue that this limit is also responsible for the residual coherent behavior of the magnetization at $\alpha=1/2$ observed in Fig.~\ref{fig:sigmazt} for long times.
    
    A similar analysis of the magnetization dynamics can be performed in the biased case, considering both the well-known limits of weak coupling and Toulouse point (see respectively App.~\ref{app:Biased} and \ref{app:Toul}). As expected, in the weak coupling regime an excellent agreement can be found with the analytical curves, while in the strong coupling limit the observed deviations from the analytic results can be traced back to the same reason as in the unbiased case.   
	
\begin{figure}[tb]
			\centering
			\includegraphics[width=\linewidth]{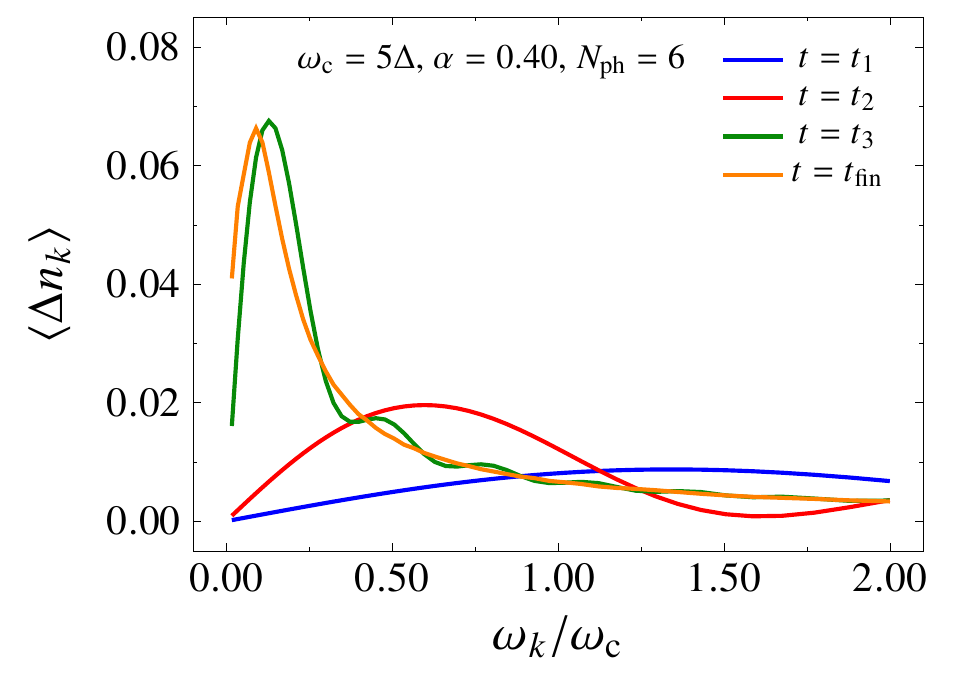}
			\caption{Expectation values of $\Delta n_{k}(t)$ computed for each bath mode $k$ at different times $\set{t\ped{1},t\ped{2},t\ped{3},t\ped{f}}=\Set{0.03,0.06,1.14,6.00}$ (in units of $\Delta\ped{r}^{-1}$), for fixed coupling strength $\alpha=0.40$, $\omega\ped{c}=5\Delta$, $T=0$, $h_0=0$, $M=50$ and $N\ped{ph}=6$.}
			\label{fig:phonons}
	\end{figure}

\begin{figure}[tb]
		\centering
		\includegraphics[width=\linewidth]{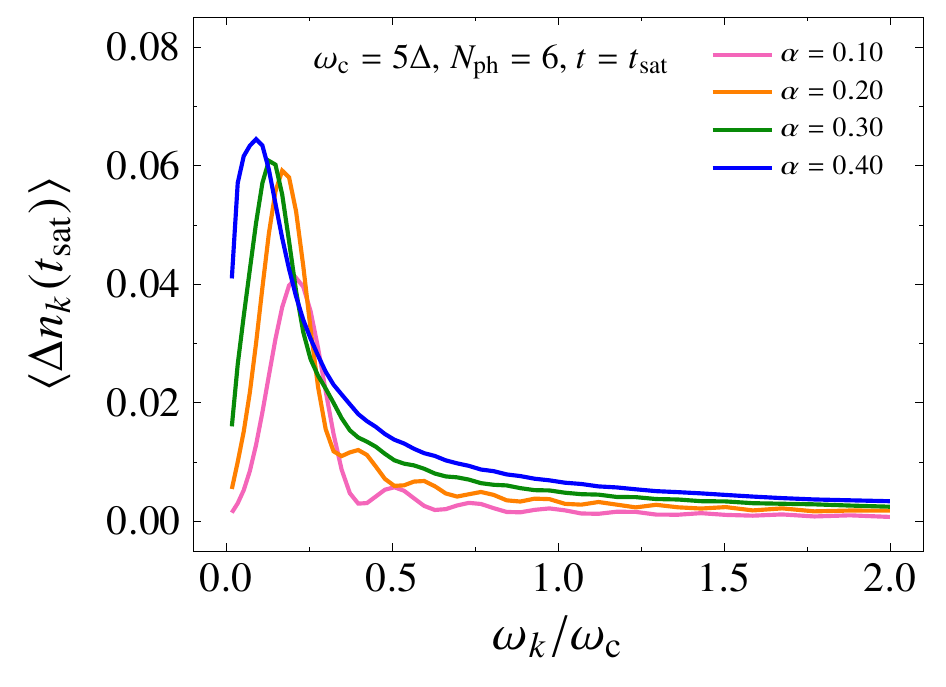}
		\caption{Expectation values of $\ev{\Delta n_{k}}$ computed for each bath mode $k$ at rescaled time $t\ped{sat}=4.85$ (in units of $\Delta\ped{r}^{-1}$), for different coupling strengths $\alpha$ in the range $\numrange[range-phrase = \text{~to~},exponent-product=\cdot]{1e-1}{4e-1}$, $\omega\ped{c}=5\Delta$, $T=0$, $h_0=0$, $M=50$ and $N\ped{ph}=6$.}
		\label{fig:phonons2}
	\end{figure}

	Additional insights can be derived from the analysis of the expectation values of the difference of number operators from their initial equilibrium values, 
	\ie~$\ev{\Delta n_{k}(t)}=\ev{n_{k}(t)}-n\api{0}\ped{k}$, computed for each bosonic mode at fixed time intervals in the range $[t\ped{0}, t\ped{f}]$, as shown in Fig.~\ref{fig:phonons};              
	it can be inferred that at short times the bath response extends over the whole frequency spectrum, high-frequency modes showing slightly greater occupation than the slower ones, even if the occupation is quite small. At intermediate times, a set of peaks start to come into play, due to multiple scattering processes of the qubit with the bath modes.
	The bath response shows a first order peak which signals the onset of a resonant behavior, its position shifting towards lower frequencies as long as time increases. 
	As expected, the behavior of each curve at intermediate times, as well as at longer times shows a clear dependence on the value of the coupling strength $\alpha$. In Fig.~\ref{fig:phonons2}, we plot $\ev{\Delta n_{k}(t)}$, computed at sufficiently long time $t\ped{sat}$ for different coupling strengths $\alpha$: the results show that 
	the position of the first-order peak shifts towards lower energies for increasing coupling strengths, and the characteristic energy of the system is proportional to the effective tunnelling energy $\Delta\ped{eff}$. Moreover, the curves of bosonic excitations exhibit oscillations in $\omega_k/\omega\ped{c}$ that tend to disappear as the coupling strength approaches the crossover value: this effect can be seen as a consequence of the increasingly incoherent behavior of the system. These features confirm that, for coupling strengths in the range $0\leq\alpha \leq 0.5$ the dynamical evolution of the whole system reaches an equilibrium state that can be interpreted in terms of a single qubit whose tunnelling energy is renormalized proportionally to $\Delta\ped{r}$, experiencing incoherent tunnelling between localized states.
	
\begin{figure}[tb]
		\centering
		\includegraphics[width=\linewidth]{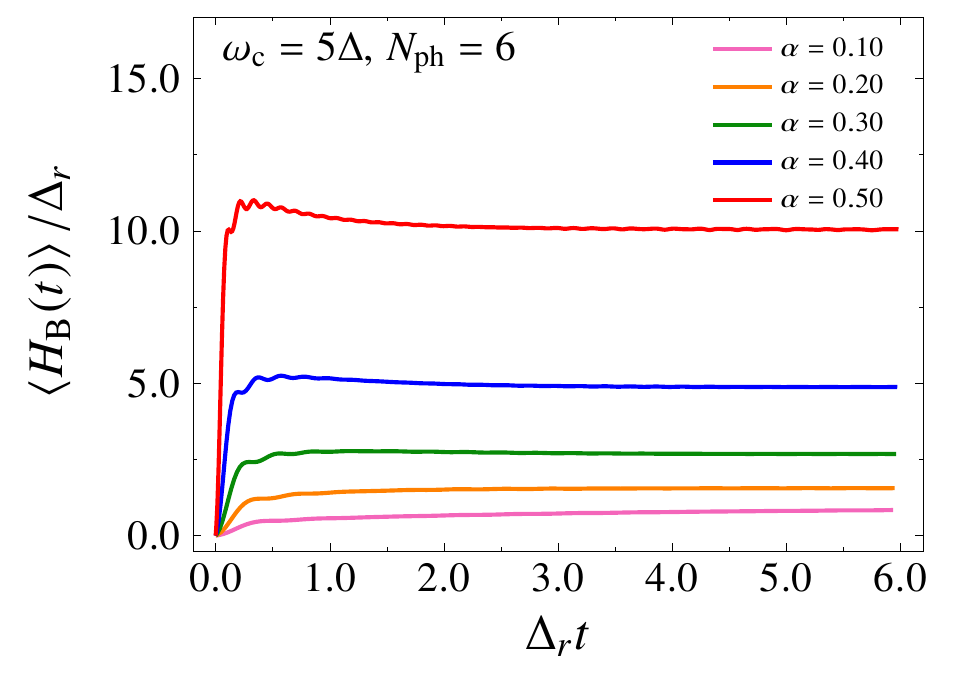}
		\caption{Expectation values of $\ev*{H\ped{B}(t)}$ (in units of $\Delta\ped{r}$ ) computed as a function of the rescaled time  $\Delta\ped{r} t$, for different coupling strengths $\alpha$ in the range $\numrange[range-phrase = \text{~to~},exponent-product=\cdot]{1e-1}{5e-1}$, $\omega\ped{c}=5\Delta$, $T=0$, $h_0=0$, $M=50$ and $N\ped{ph}=6$.}
		\label{fig:bathenergy}
	\end{figure}

	The exchanged energy with the bath can also be studied for different values of the coupling strengths $\alpha$: from Fig.~\ref{fig:bathenergy}, it can be shown that all the curves tend to a saturation value, which in our simulation is fixed by the energy conservation: as shown in \cite{Cangemi:SIL}, at every time $t$ the energy of the non-equilibrium initial state of the qubit is equal to the sum of the expectation values of the different operators in Eq.~\eqref{eq:TotalHam}. It can also be noticed that the saturation value of $\ev*{H\ped{B}(t)}$ strongly depends on the coupling strength $\alpha$. Moreover, for increasing coupling strengths, it can be observed that the bath energy exhibits an oscillatory behavior at short times, and a moderately prononunced peak which tends to be greater than its long-time value. This feature is due to the increasing importance of the qubit-bath correlations which, in the strong coupling regime, can modify the mechanism of energy exchange.        
	
	\section{Time dependent protocol}\label{sec:Magnetic} 
    
    In this section, we study the effect of decoherence and dissipation on a two level system  subject to a rotating magnetic field. We take the qubit Hamiltonian as in Eq.~\eqref{eq:QubitHam}, where $h$ is the magnitude of the applied magnetic field; we adopt a system of polar coordinates $(\theta,\phi)$, \ie~$\boldsymbol{h}=(h\sin\theta\cos\phi,h\sin\theta\sin\phi, h\cos\theta)$. We also introduce an additional static magnetic field along the positive $\hat{z}$ direction, \ie~$h\ped{0}\hat{z}$. We restrict the rotating magnetic field $\boldsymbol{h}$ in the $\hat{x}\mbox{-}\hat{z}$ plane by fixing $\phi=0$. The qubit Hamiltonian thus reads 
    \begin{equation}\label{eq:qubitperiodic}
    H\ped{S}(t)= -\frac{1}{2}(h\ped{0}+h\cos\theta(t))\sigma_z - \frac{h}{2}\sin\theta(t) \sigma_x  
    \end{equation}  
    The magnetic field $\boldsymbol{h}$ evolves performing a sweep in $\hat{x}\mbox{-}\hat{z}$ plane in a total time $t\ped{f}$, \ie~the polar angle changes according to $\theta\left(t\right)=\pi(t-t\ped{0})/t\ped{f}$, from $\theta(t\ped{0})=0$ to $\theta(t\ped{f})=\pi(1-t\ped{0}/t\ped{f})$. 
    This protocol, widely studied in the field of Nuclear Magnetic Resonance (NMR), has regained attention following recent theoretical and experimental works \cite{Gritsev,RoushanNature:topotrans}. It has been shown that physical implementations of Hamiltonians of the form of Eq.~\eqref{eq:qubitperiodic} can be achieved with high level of control employing superconducting circuits; 
    moreover, a simple mapping exists from Eq.~\eqref{eq:qubitperiodic} to the Haldane model at half filling on a honeycomb lattice \cite{Haldane88}, which is a prototypical model of a Chern insulator. Following this mapping, every qubit state on the Bloch sphere at fixed coordinates $(\theta,\phi)$ can be mapped onto a single quasi-momentum state $(k\ped{x},k\ped{y})$ around the high-symmetry points of the first Brillouin zone of the honeycomb lattice. As a consequence, it allows for a dynamical measurement of the topological properties of the Haldane model by making use of a superconducting qubit, \eg~the first Chern number can be probed. It follows that, by tuning the ratio of the field amplitudes $h\ped{0}/h$ and performing quantum state tomography  at different times $t$ during the sweep, topological transitions can be measured with high level of accuracy.        
    
    In the following, we analyze the dissipative dynamics of a qubit described by Eq.~\eqref{eq:qubitperiodic} at weak and strong coupling strengths, both for long and short sweep times $t\ped{f}$ as compared with the time scale $1/h$, \ie~we consider both adiabatic and anti-adiabatic regimes. The qubit is coupled to the environment along a direction which lies in the plane of rotation of the magnetic field, and we focus on the two particular cases $\boldsymbol{\hat{n}}=\hat{z}\mbox{, }\hat{x}$. 
    We compute the excess energy of the qubit at the end of the sweep, \ie~the difference between the mean value of the reduced system energy and the ground state energy $\epsilon_{gs}(t\ped{f})$ of the non-interacting qubit Hamiltonian in Eq.~\eqref{eq:qubitperiodic}, computed at final time $t\ped{f}$    
    \begin{equation}
    \epsilon\ped{res}=\Tr[\rho(t\ped{f})H\ped{S}(t\ped{f})]-\epsilon_{gs}(t\ped{f})   
    \end{equation}
    Due to the simple form of Eq.~\eqref{eq:qubitperiodic}, the excess energy can also be linked to the fidelity $\mathcal{F}(t\ped{f})$ at the end of the sweep
    \begin{equation}
    \epsilon\ped{res}=\abs{h-h\ped{0}}(1-\mathcal{F}(t\ped{f}))   
    \end{equation}  
     where $\mathcal{F}(t\ped{f})=\bra*{\psi\ped{gs}(t\ped{f})}\rho\ped{S}(t\ped{f})\ket*{\psi\ped{gs}(t\ped{f})}$ and $\ket*{\psi\ped{gs}(t\ped{f})}$ is the ground state of qubit Hamiltonian in Eq.~\eqref{eq:qubitperiodic} at $t=t\ped{f}$. In addition, we compute the expectation values of qubit operators $\ev{\boldsymbol{\sigma}}=(\ev{\sigma_x(t)}, \ev*{\sigma_y(t)}, \ev{\sigma_z(t)})$ as functions of time, \ie~the dynamical evolution of the Bloch vector, at fixed final times $t\ped{f}$ and for different values of the coupling strength.    
    
    We first consider the qubit system in the absence of dissipation, taking the static bias field $h\ped{0}=0$: at initial time $t_0=0$, the magnetic field is aligned along the positive $\hat{z}$ direction and the qubit is prepared in its ground state, \ie~$\ket{\psi(t\ped{0})}=\ket{\hat{z},+}$. For $t>0$, the field $\boldsymbol{h}$ rotates around the $\hat{y}$ axis. The qubit dynamics can be straightforwardly solved in the counter-rotating frame around the $\hat{y}$ axis (see App.~\ref{app:unitary}): due to its simple form, the Hamiltonian in Eq.~\eqref{eq:qubitperiodic} in the rotating frame is time-independent, and it follows that the excess energy of the closed system reads 
    \begin{equation}\label{eq:eresclosed}
    \epsilon\ped{res} =\frac{h \dot{\theta}^2}{2} \frac{1-\cos(\pi\sqrt{h^2 + \dot{\theta}^2}/\dot{\theta})}{h^2+ \dot{\theta}^2}
    \end{equation}
    where we put for brevity $\dot{\theta}=\pi/t\ped{f}$. The qubit dynamics is described by a cycloid on the Bloch sphere, \ie~the Bloch vector periodically points out of the $\hat{x}\mbox{-}\hat{z}$ plane. This trajectory is due to the oscillations in time of the magnetization along the $\hat{y}$ axis: therefore, $\ev*{\sigma_y(t)}$ can serve as a measure of deviation from the adiabatic path, which is a circle in the $\hat{x}\mbox{-}\hat{z}$ plane. The non-adiabatic response of the Bloch vector is thus proportional to $\ev*{\sigma_y(t)}$ \cite{Gritsev}; furthermore, using perturbation theory it has been shown that at first order in $\dot{\theta}/h$ the non-adiabatic response can be linked to the curvature of the ground state manifold of the Hamiltonian in Eq.~\eqref{eq:qubitperiodic}, \ie~to the Berry phase of the qubit. As a consequence, the measure of  $\ev*{\sigma_y(t)}$ at each time $t$ allows to achieve the fidelity at final time $t\ped{f}$ that, in the quasi-adiabatic limit, 
    can be used to compute the first Chern number of the system \cite{RoushanNature:topotrans}.
    The deviation from the adiabatic path can also be seen from the excess energy in Eq.~\eqref{eq:eresclosed}, which is plotted in Fig.~\ref{fig:eressigmaz} (black curve): notice that it exhibits several maxima corresponding to different final times $t\ped{f}$ owing to the fact that, in the non-adiabatic regime, the qubit dynamics cannot follow the evolution of the externally driven magnetic field, and the state vector of the qubit at the end of the sweep differs from the corresponding ground state. However, the amplitude of these maxima is decreasing as long as the time $t\ped{f}$ is increased, \ie~the dynamics can be considered truly adiabatic only in the limit $t\ped{f}\gg1/h$. 
   
    This scenario undergoes several changes if the interaction with the external bath is considered. 
    \subsection{Coupling along $\hat{z}$}\label{subsec:couplingz}
\begin{figure}[tb]
    	\centering
    	\includegraphics[width=\linewidth]{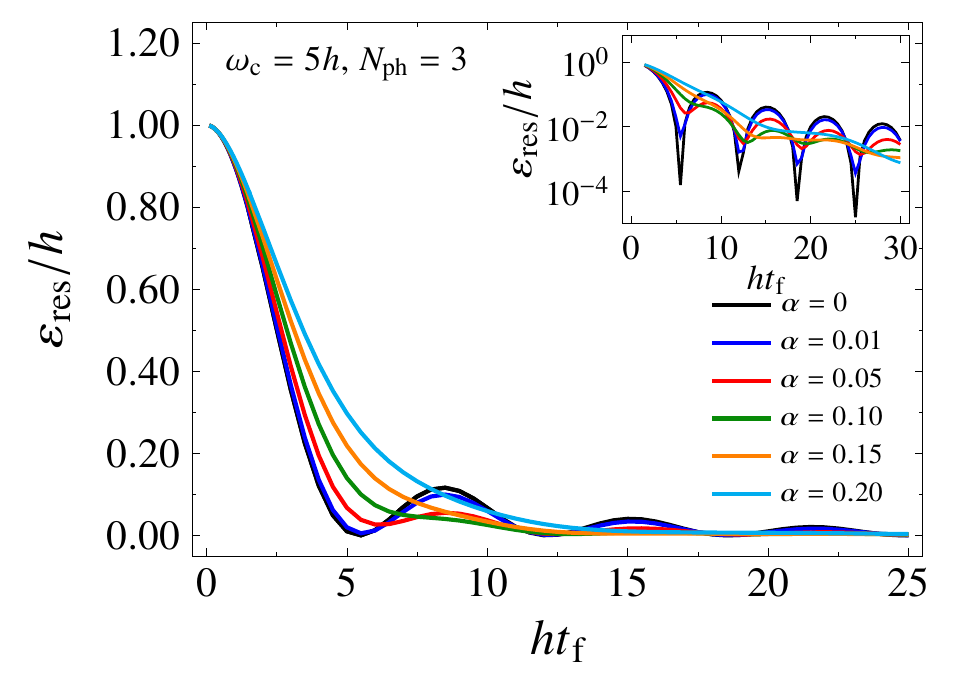}
    	\caption{Excess energy plotted as a function of the final time $t\ped{f}$ (in units of $h^{-1}$), for different coupling strengths ranging from $\numrange[range-phrase = \text{~to~},exponent-product=\cdot]{1e-2}{2e-1}$, in the case $\boldsymbol{\hat{n}}=\hat{z}$. The number of modes has been fixed to $M=80$, the cutoff frequency $\omega\ped{c}=5 h$, $N\ped{ph}=3$ and $T=0$. Inset: semi-logarithmic plot of the same curves as in the main plot.}
    	\label{fig:eressigmaz}
    \end{figure}

    We first analyze the case of interaction along $\hat{z}$ axis. In Fig.~\ref{fig:eressigmaz}, we plot the excess energy curve of the qubit interacting with a bath at $T=0$, for different values of the coupling strength $\alpha$ ranging from $\numrange[range-phrase = \text{~to~},exponent-product=\cdot]{0}{2e-1}$. 
    As it can be noticed, the interaction with the external bath acts to reduce the coherence of the system dynamics. The effect of decoherence results in a smoothing of the excess energy curve with respect to the closed case. However, the difference between the closed and the open system curve depends on the final time $t\ped{f}$: at short final times $t\ped{f}$, the interaction with the environment generally leads to an increase of the residual energy, resulting in a non-adiabatic behavior of qubit dynamics; conversely, at intermediate times $t\ped{f}$, the effect of the bath can lead to a decrease of the local maxima of the excess energy as compared with the closed case, \ie~the state of the qubit at $\theta(t\ped{f})$ is closer to the corresponding state on the adiabatic path. It follows that, at weak coupling regime the effect of friction counteracts the non-adiabaticity of the system induced by the fast external drive, thus resulting in a reduction of the excess energy. This scenario changes in the intermediate coupling regime: for coupling strengths $\alpha > 0.1$, it can be noticed that the excess energy starts to increase, and the system definitely misses the adiabatic path. The resulting non-monotonic behavior of the excess energy can be clearly observed for final times $t\ped{f}$ where the closed system curve shows secondary maxima of excitation, while for values of $t\ped{f}$ corresponding to minima the interaction with the bath leads to monotonic non-adiabaticity. It should also be noticed that, as depicted in the inset of Fig.~\ref{fig:eressigmaz}, for very slow sweeps the open system curves at weak coupling strengths tend to coincide, and they are consistent with the closed system result. Interestingly, the monotonic non-adiabaticity at strong coupling regime was recently observed in \cite{henriet:Topologysbm}, where the dynamical behavior of the Chern number in a dissipative environment was studied and a description of the bath-induced non-adiabaticity was achieved using non-perturbative Stochastic Schr\"odinger equation. 
    
\begin{figure}[tb]
    	\centering
    	\includegraphics[width=\linewidth]{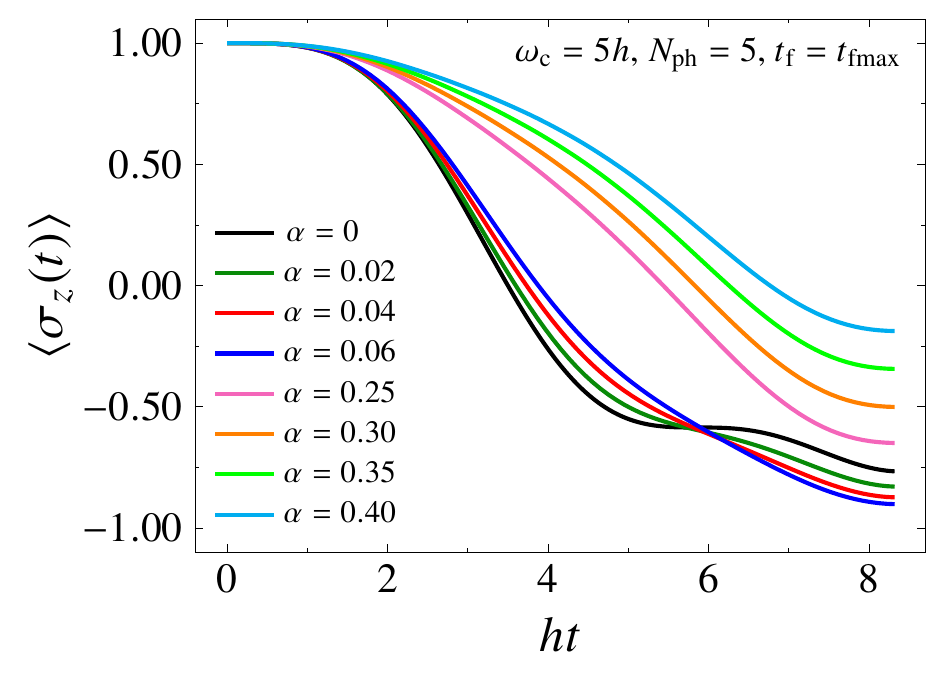}
    	\caption{Plot of $\ev{\sigma_z(t)}$ as a function of time $t$ for the protocol in Eq.~\eqref{eq:qubitperiodic}, with fixed final time $t\ped{f}=t\ped{fmax}=8.42/h$. The qubit couples to an Ohmic bath ($s=1$) along $\boldsymbol{\hat{n}}=\hat{z}$ , for different coupling strengths ranging from $\numrange[range-phrase = \text{~to~},exponent-product=\cdot]{0}{4e-1}$. The number of modes has been fixed to $M=70$, the cutoff frequency $\omega\ped{c}=5 h$, $N\ped{ph}=5$ and $T=0$.}
    	\label{fig:evolutionsigmaz}
    \end{figure}

\begin{figure}[tb]
    	\centering
    	\includegraphics[width=\linewidth]{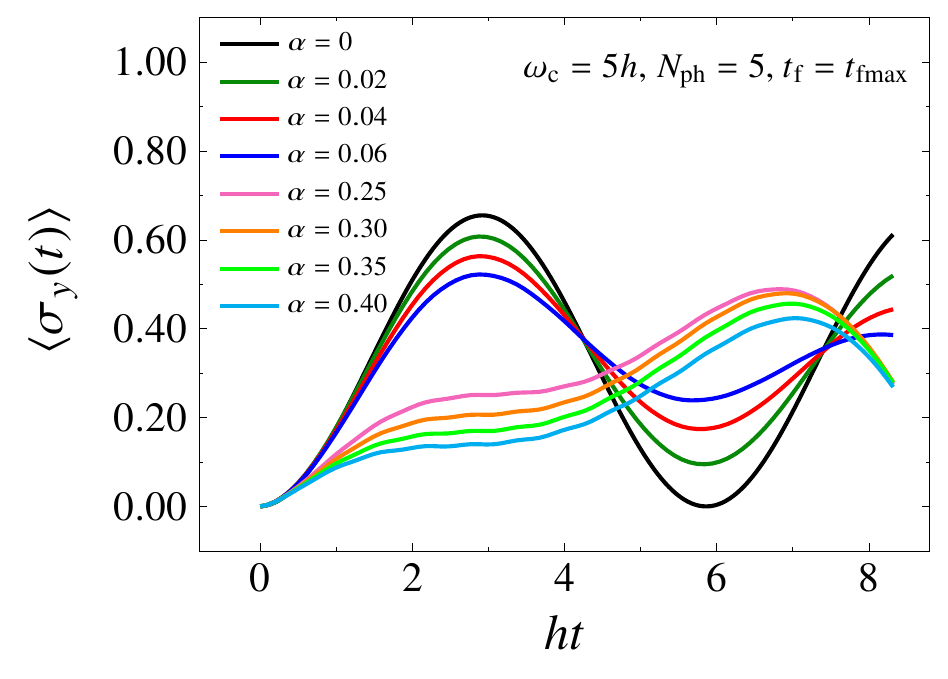}
    	\caption{Plot of $\ev*{\sigma_y(t)}$ as a function of time $t$ for the protocol in Eq.~\eqref{eq:qubitperiodic}, with fixed final time $t\ped{f}=t\ped{fmax}=8.42/h$. The qubit couples to an Ohmic bath ($s=1$) along $\boldsymbol{\hat{n}}=\hat{z}$, for different coupling strengths ranging from $\numrange[range-phrase = \text{~to~},exponent-product=\cdot]{0}{4e-1}$. The number of modes has been fixed to $M=70$, the cutoff frequency $\omega\ped{c}=5 h$, $N\ped{ph}=5$ and $T=0$.}
    	\label{fig:evolutionsigmay}
    \end{figure}

\begin{figure}[tb]
    	\centering
    	\includegraphics[width=\linewidth]{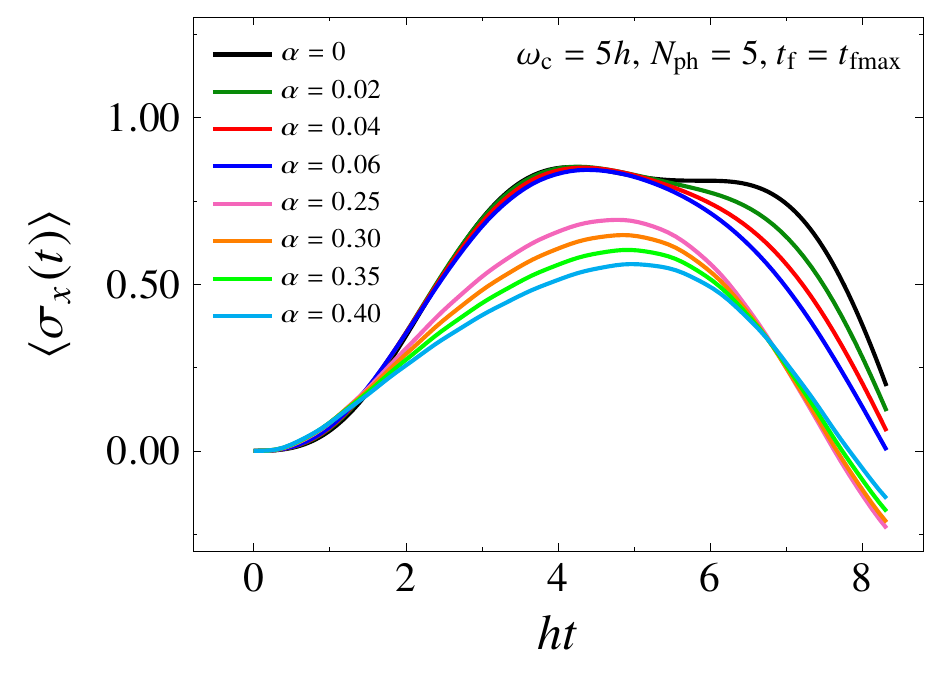}
    	\caption{Plot of $\ev{\sigma_x(t)}$ as a function of time $t$ for the protocol in Eq.~\eqref{eq:qubitperiodic}, with fixed final time $t\ped{f}=t\ped{fmax}=8.42/h$. The qubit couples to an Ohmic bath ($s=1$) along $\boldsymbol{\hat{n}}=\hat{z}$, for different coupling strengths ranging from $\numrange[range-phrase = \text{~to~},exponent-product=\cdot]{0}{4e-1}$. The number of modes has been fixed to $M=70$, the cutoff frequency $\omega\ped{c}=5 h$, $N\ped{ph}=5$ and $T=0$.}
    	\label{fig:evolutionsigmax}
    \end{figure}

    Further information on the dynamics of the open system at intermediate final times $t\ped{f}$ can be derived from the analysis of the expectation values $\ev{\sigma_x(t)},\ev*{\sigma_y(t)},\ev{\sigma_z(t)}$: in Fig.~\ref{fig:evolutionsigmaz}, \ref{fig:evolutionsigmay}, \ref{fig:evolutionsigmax},
    we plot the expectation values of the spin operators as a function of time $t$, from weak to strong coupling regime and for fixed final time $t\ped{f}=t\ped{fmax}=8.42/h$, corresponding to the first second-order maximum of Eq.~\eqref{eq:eresclosed}. It can be noticed that, following the Heisenberg equations which link the time derivative of $\ev{\sigma_z(t)}$ to $\ev*{\sigma_y(t)}$, the decrease in the excess energy occurring for $t\ped{f}=t\ped{fmax}=8.42/h$ observed at weak coupling can be traced back to the progressive change of $\ev*{\sigma_y(t)}$. 
    
    Hence, for increasing coupling strengths $\alpha$ it can be noticed that the magnetization along $\hat{y}$ loses the oscillatory behavior with frequency $\sqrt{\dot{\theta}^2 + h^2}$ (see App.~\ref{app:unitary}), which is a characteristic feature of dynamics in the absence of dissipation: actually, as it can be inferred from Fig.~\ref{fig:evolutionsigmay} the second local minimum turns into a local maximum, its position drifts towards higher times $t$, causing the inflection point of $\ev{\sigma_z(t)}$ in Fig.~\ref{fig:evolutionsigmaz} to change accordingly; eventually, at the end of the sweep the magnetization along $\hat{z}$ tends to the adiabatic value, and the state of the qubit in the open dynamics at final time $t\ped{f}$ is closer to the ground state $\ket*{\psi\ped{gs}(t\ped{f})}$. 
    While the previous description holds true also when the closed-system excess energy in Eq.~\eqref{eq:eresclosed} shows local minimum values, \eg~$t\ped{f}=t\ped{fmin}=12.7/h$, it can be observed that at weak coupling strengths the interaction with the bath cannot noticeably change the excess energy.      
    
     More information on the physics at strong coupling regime can be drawn:
     as shown in Fig.~\ref{fig:evolutionsigmaz}, for $\alpha > 0.10$  $\ev{\sigma_z(t\ped{f})}$ starts to increase. This behavior clearly depends on the final time $t\ped{f}$, \ie~on the slope of the external drive: for faster sweeps, the non-adiabatic behavior due to the interaction with the environment occurs at lower coupling strengths as compared to slower evolutions;
     \begin{figure}[thb]
     	\centering
     	\includegraphics[width=\linewidth]{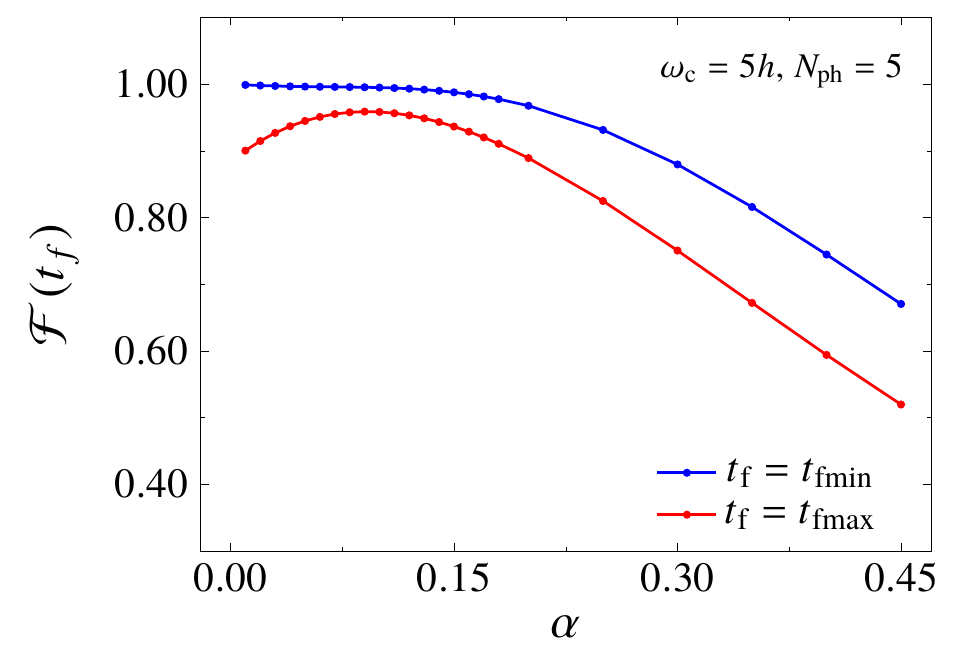}  	
     	\caption{Fidelity at final time $t\ped{f}$ $\mathcal{F}(t\ped{f})$, plotted against the coupling strength $\alpha$ in the range $\numrange[range-phrase = \text{~to~},exponent-product=\cdot]{1.0e-2}{4.5e-1}$, for two fixed final times $t\ped{f}=\set{t\ped{fmax},t\ped{fmin}}=\set{8.42/h,12.17/h}$ corresponding to the first second order maximum and the second minimum of Eq.~\eqref{eq:eresclosed}. The number of modes has been fixed to $M=70$, the cutoff frequency $\omega\ped{c}=5 h$, $N\ped{ph}=5$ and $T=0$.} 	
     	\label{fig:fidelity}       
     \end{figure}
     as a consequence, the coupling strength directly influences the adiabaticity condition. 
     This feature can be inferred from Fig.~\ref{fig:fidelity}, where we plot the behavior of fidelity $\mathcal{F}(t\ped{f})$ at the end of the sweep, computed for two different fixed final times $t\ped{f}=\set{t\ped{fmax},t\ped{fmin}}=\set{8.42/h,12.17/h}$, corresponding to the first second-order maximum and the second minimum of Eq.~\eqref{eq:eresclosed} (see Fig.~\ref{fig:eressigmaz}, black curve), for different coupling strengths $\alpha$ taken in the range $\numrange[range-phrase = \text{~to~},exponent-product=\cdot]{1.0e-1}{4.5e-1}$. At final time $t\ped{f}=t\ped{fmax}$, where the closed system excess energy exhibits a local maximum, fidelity shows a small non-monotonic behavior, due to the previously described effect; conversely, at $t\ped{f}=t\ped{fmin}$ a flat behavior at weak coupling, followed by monotonic decrease occurring at higher values of $\alpha$ can be observed.

   As shown in \cite{henriet:Topologysbm}, an adiabaticity criterion for the protocol in Eq.~\eqref{eq:qubitperiodic} has been proposed which links the velocity of the sweep $\dot{\theta}$ to the renormalized field $\Delta\ped{r}$ along $\hat{x}$ direction (with $\Delta=h$), \ie~$\dot{\theta}\ll \Delta\ped{r}$, provided that $\dot{\theta}\ll h$. Further, for fixed values of $\dot{\theta}$ well below $h$, at strong coupling a crossover from quasi-adiabatic to non-adiabatic behavior occurs at $\dot{\theta}\simeq \Delta\ped{r}$. We find that our numerical results at strong coupling generally agree with this scenario, while in the weak coupling regime several intervals of final times $t\ped{f}$ exist where the bath can act to improve the adiabaticity. It follows that, at weak coupling strengths the dynamical measure of the topological properties shows robustness to the external noise.

    \subsection{Coupling along $\hat{x}$}\label{subsec:couplingx}
     
\begin{figure}[tb]
     	\centering
     	\includegraphics[width=\linewidth]{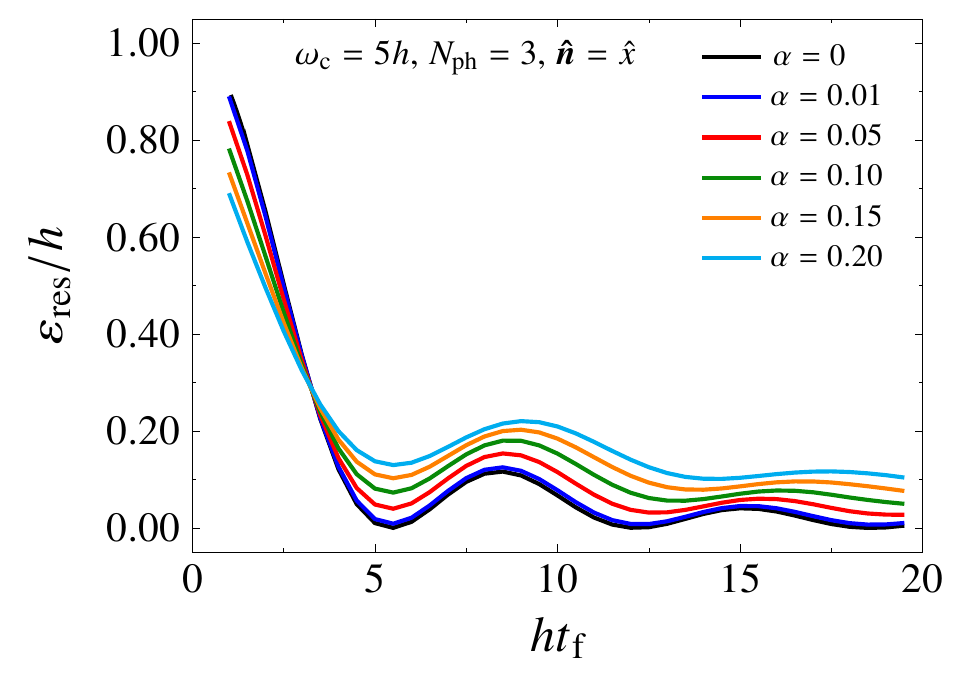}
     	\caption{Excess energy plotted as a function of the final time $t\ped{f}$ (in units of $h^{-1}$), for different coupling strengths $\alpha$ ranging from $\numrange[range-phrase = \text{~to~},exponent-product=\cdot]{1e-2}{2e-1}$, in the case of $\boldsymbol{\hat{n}}=\hat{x}$. The number of modes has been fixed to $M=80$, the cutoff frequency $\omega\ped{c}=5 h$, $N\ped{ph}=3$ and $T=0$.}
     	\label{fig:eres-sigmax}
     \end{figure}

     Qualitatively different results can be found if the qubit couples with the bath along $\hat{x}$ axis. Here we restrict to weak coupling regime and simulate the dissipative dynamics at $T=0$ of the time-dependent protocol in Eq.~\eqref{eq:qubitperiodic} with $\boldsymbol{\hat{n}}=\hat{x}$ and $h\ped{0}=0$. In Fig.~\ref{fig:eres-sigmax}, we plot the excess energy as a function of the final time $t\ped{f}$ for different coupling strengths, taken in the same range as in Fig.~\ref{fig:eressigmaz}. 
   
\begin{figure}[tb]
	  	\centering
	  	\includegraphics[width=\linewidth]{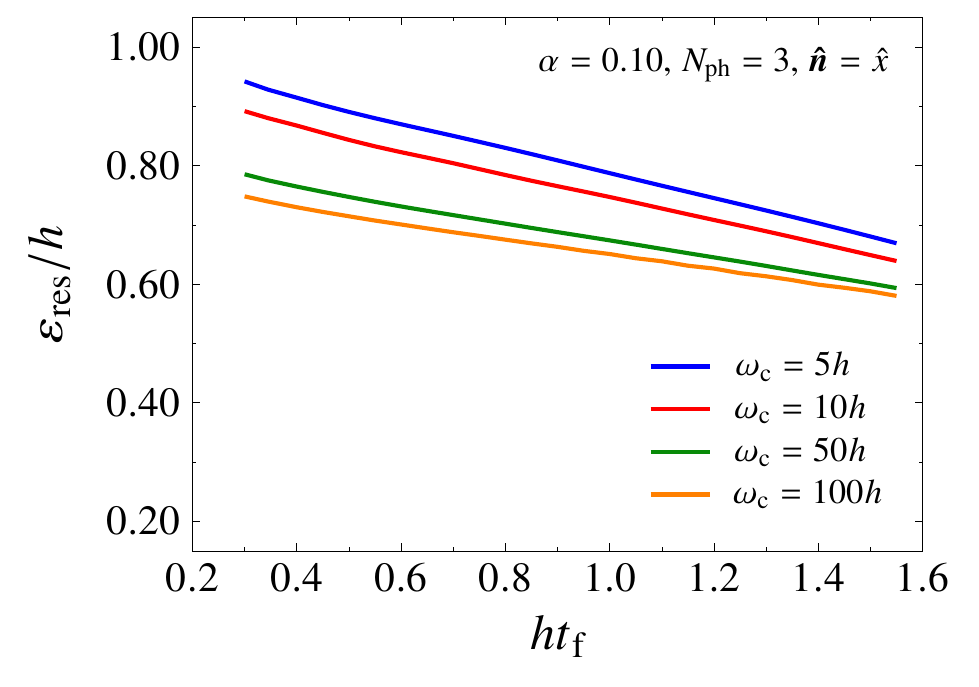}
	  	\caption{Excess energy plotted as a function of the final time $t\ped{f}$ chosen in the range $\numrange[range-phrase = \text{~to~},exponent-product=\cdot]{0.2}{1.5}$ (in units of $h^{-1}$), for fixed $\alpha=0.10$, $M=80$, $N\ped{ph}=3$, $T=0$ and different cutoff frequencies $\omega\ped{c}$.}
	  	\label{fig:eres-sigmax-highomega}
	  \end{figure}

     It can be shown that, for very fast sweeps the excess energy can be lower than the closed system result: the actual numerical results at short final times $t\ped{f}$ depend on the coupling strength $\alpha$ at fixed cutoff frequency $\omega\ped{c}$. As shown in Fig.~\ref{fig:eres-sigmax-highomega}, by increasing the cutoff frequency $\omega\ped{c}$, the short final time limit of the excess energy curve decreases, as a result of the reduced reaction time of the bath. However, the choice of different cutoff frequencies $\omega\ped{c}$ does not qualitatively change the physics at long times $t\ped{f}$.         
	 The decrease in the excess energy at short $t\ped{f}$ is due to the peculiar form of the coupling to the external environment, which causes the qubit to flip at a fixed rate proportional to the coupling strength. This effect can provide a slight advantage to the success of the protocol, as long as the final time $t\ped{f}$ is sufficiently short. However, for longer final times $t\ped{f}$ the open system excess energy tends to be greater than the closed curve: this effect leads to an increasingly non-adiabatic dynamics, even at weak coupling strengths, as opposed to the case studied in \ref{subsec:couplingz} where, as long as the closed system dynamics is quasi-adiabatic, the dynamics is unaffected by the environment. This result shows several analogies with a recent study of the finite-time LMSZ protocol \cite{SunfinitetimeLandZen}, showing that the effect of a transverse coupling to the bath at long final times $t\ped{f}$ can lead to a fidelity $\mathcal{F}(t\ped{f})$ lower than $1$, as opposed to the exact result proposed in \cite{wubs:lz-zero-t}. In addition, here the effect of time-periodic driving can be clearly observed, noticing the persistence of a structure made of several secondary maxima in the excess energy. These findings point towards an increasingly non-adiabatic behavior due to the bath, as long as the coupling strengths increases, occurring at intermediate up to long final times $t\ped{f}$.

    \section{Conclusions}\label{sec:conclusions}
	In this work, we studied the dynamics of a qubit in contact with its environment, subject both to static and driven external fields, from weak to strong coupling strengths, using the SIL approach. We showed that our method can provide a good description of the physics of the SBM as a function of the coupling strength up to Toulouse point, where a crossover from coherent to incoherent behavior of the qubit magnetization takes place. We provided additional insights on the dynamics of the bath degrees of freedom, showing the changes in the bath response as a function of time. Moreover, we studied a protocol of a driven qubit subject to a time-periodic driving, with dissipation taking place along different directions. We showed that in the case of coupling along $\hat{z}$, if the dissipation strength is sufficiently weak, the influence of the bath can counteract the non-adiabaticity of the closed system evolution, leading to a non-monotonic behavior of the fidelity as a function of the coupling strength at fixed values of the final times. Conversely, at strong coupling bath-induced non adiabaticity \cite{henriet:Topologysbm} takes place, hindering the success of the protocol. This scenario changes if the coupling along $\hat{x}$ axis is considered: a measurable advantage over the closed system dynamics can be observed only for very fast sweeps, while for longer sweep durations we predict an increasingly non-adiabatic behavior, \ie~the excess energy tends to increase at increasing coupling strength. In the near future, we plan to extend our analysis to recently proposed time-dependent protocols implementing counter-diabatic driving \cite{Polkovnikov:CD,SelsE3909}, in order to investigate the influence of the environment on the final success probability of these protocols in a broad range of coupling regimes. In addition, energy exchanges between systems of externally driven interacting qubits and the bath will also be analyzed, as well as prototypical models of quantum heat engines.

	\appendix

	\section{The biased case - weak coupling regime}\label{app:Biased}
	
\begin{figure}[thb]
		\centering
		\includegraphics[width=\linewidth]{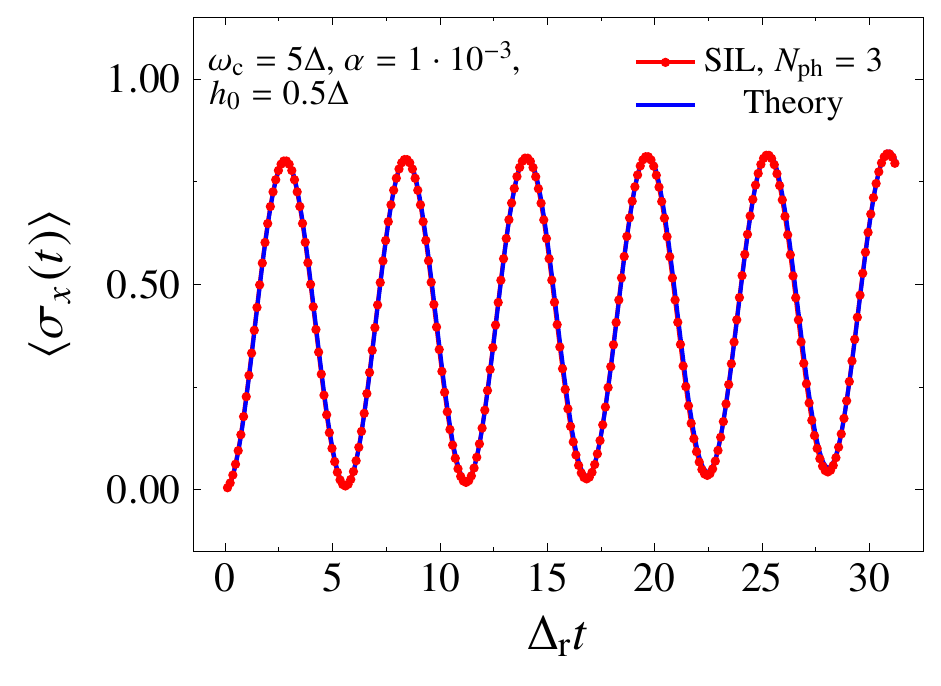}
		\caption{Plot of $\ev{\sigma_x(t)}$ as a function of time $t$ for the biased SBM, having fixed $\omega\ped{c}=5 \Delta$, $\alpha=0.001$, $h_0=0.5 \Delta$, $T=0$ and $N\ped{ph}=3$. SIL results (red points) compared with theoretical curve from Eq.~\eqref{eq:theoryweak} (solid blue curve).}
		\label{fig:weakcouplingx}
	\end{figure}

	Below, we report a comparison of our numerical results for the dynamics of the qubit magnetization in the biased SBM ($h_0\neq 0$), with analytical curves derived by means of a first order expansion in the model parameters reported in \cite{Carrega15:heatexchange}. In the limit of weak coupling regime ($\alpha\ll 1$), and taking $T=0$, the qubit magnetizations along $\hat{x}$, $\hat{z}$ axes read
	\begin{equation}
	\begin{gathered}\label{eq:theoryweak}
	\ev{\sigma_z(t)}=\frac{h_0}{\Omega}(1-e^{-\gamma_r t})+\frac{h_0^2}{\Omega^2}e^{-\gamma_r t}+\frac{\Delta\ped{r}^2}{\Omega^2} \cos(\Omega t)e^{-\tilde{\gamma}_r t}\\
    \ev{\sigma_x(t)}=\frac{\Delta\ped{r}^2}{\Delta\Omega}(1-e^{-\gamma_r t}) + \frac{h_0\Delta\ped{r}^2}{\Delta\Omega^2}(e^{-\gamma_r t} -\cos(\Omega t)e^{-\tilde{\gamma}_r t})
    \end{gathered}
	\end{equation}
	where $\Omega=\sqrt{\Delta\ped{r}^2 + h_0^2}$, and the damping rates are $\gamma_r=\pi \alpha\Delta\ped{r}^2/\Omega$, $\tilde{\gamma}_r=\gamma_r/2$. 
		\begin{figure}[thb]
			\centering
			\includegraphics[width=\linewidth]{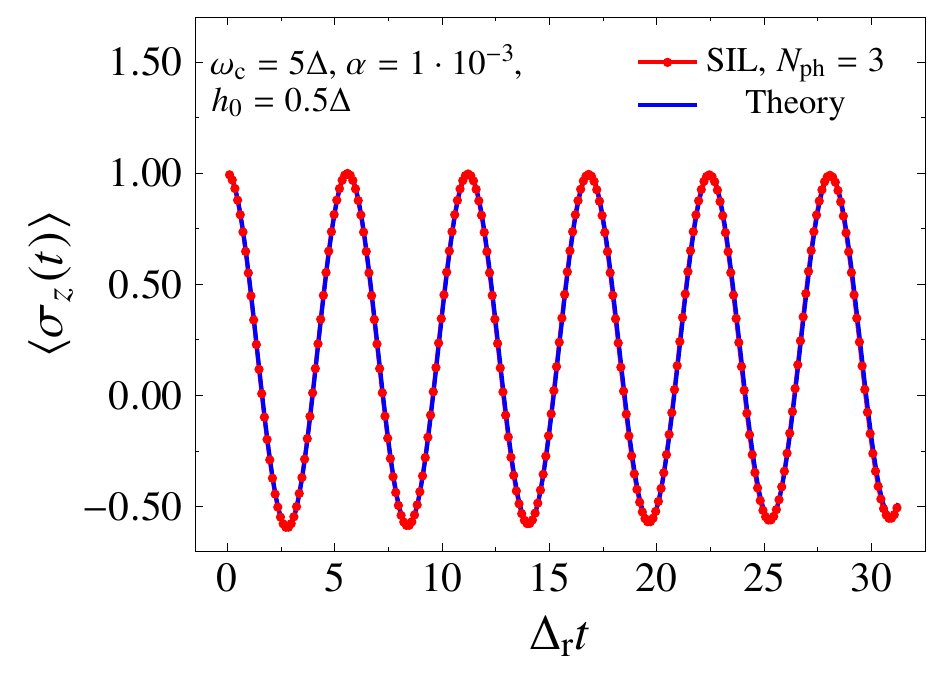}
			\caption{Plot of $\ev{\sigma_z(t)}$ as a function of time $t$ for the biased SBM, having fixed $\omega\ped{c}=5 \Delta$, $\alpha=0.001$, $h_0=0.5 \Delta$, $T=0$ and $N\ped{ph}=3$. SIL results (red points) compared with theoretical curve from Eq.~\eqref{eq:theoryweak} (solid blue curve).}
			\label{fig:weakcouplingz}
		\end{figure}
	For non-zero temperatures $T$ of the reservoir, these results slightly change 	\cite{Carrega15:heatexchange}. 
	Eq. \eqref{eq:theoryweak} include the quantum non-Markovian effects due to the interaction of the qubit with the bath.
    In Fig.~\ref{fig:weakcouplingz}, \ref{fig:weakcouplingx} we compare numerical SIL results with analytical curves in Eq.~\eqref{eq:theoryweak}, having fixed the bias field $h_0=0.5\Delta$ and the coupling strength $\alpha=1\cdot10^{-3}$.  
    As expected, SIL results show an excellent agreement with analytical curves.
    It could be shown that small quantitative differences may appear as we compare the numerical results for the energy exchanged with the reservoir with the analytical expression reported in \cite{Carrega15:heatexchange}.		
	   
	However, by means of our technique the qualitative features of the energy exchange, from intermediate to long times, can be correctly described, and we argue that the observed differences are mainly due to the choice of the small cutoff frequency $\omega_c$.   
			
	\section{Towards strong coupling regime}\label{app:Toul}
	
\begin{figure}[tbh]
     	\centering
     	\includegraphics[width=\linewidth]{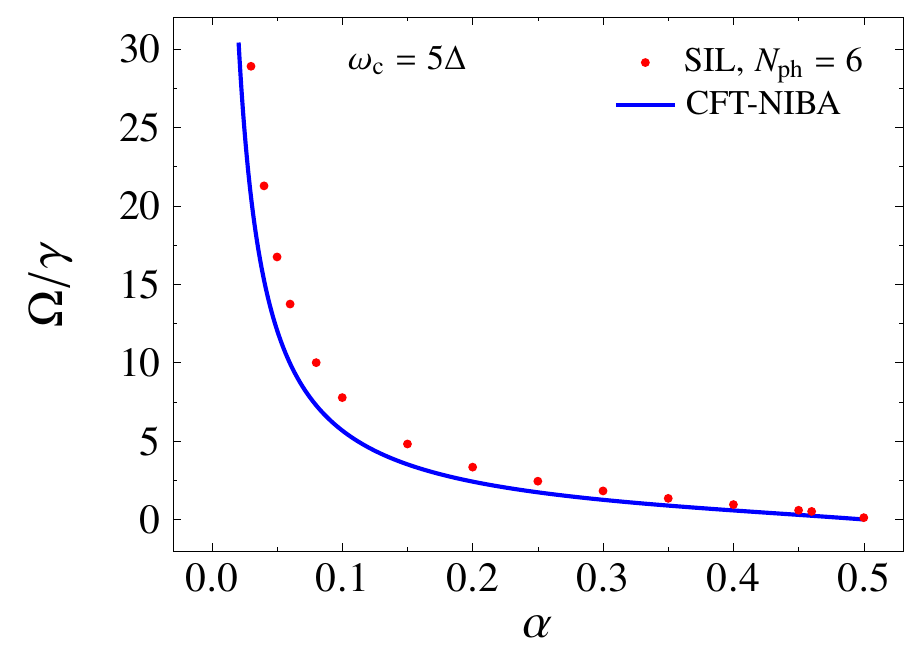}
     	\caption{Plot of the quality factor $Q$ against the coupling strength $\alpha$ at $T=0$: numerical estimate derived from a fit of the numerical curves showed in Sec.~\ref{sec:Toulouse} (red points), compared with the theoretical result in Eq.~\eqref{eq:qualityfactor} known from CFT and NIBA predictions (blue curve).}
     	\label{fig:qf}
     \end{figure}

\begin{figure}[tb]
    	\centering
		\includegraphics[width=\linewidth]{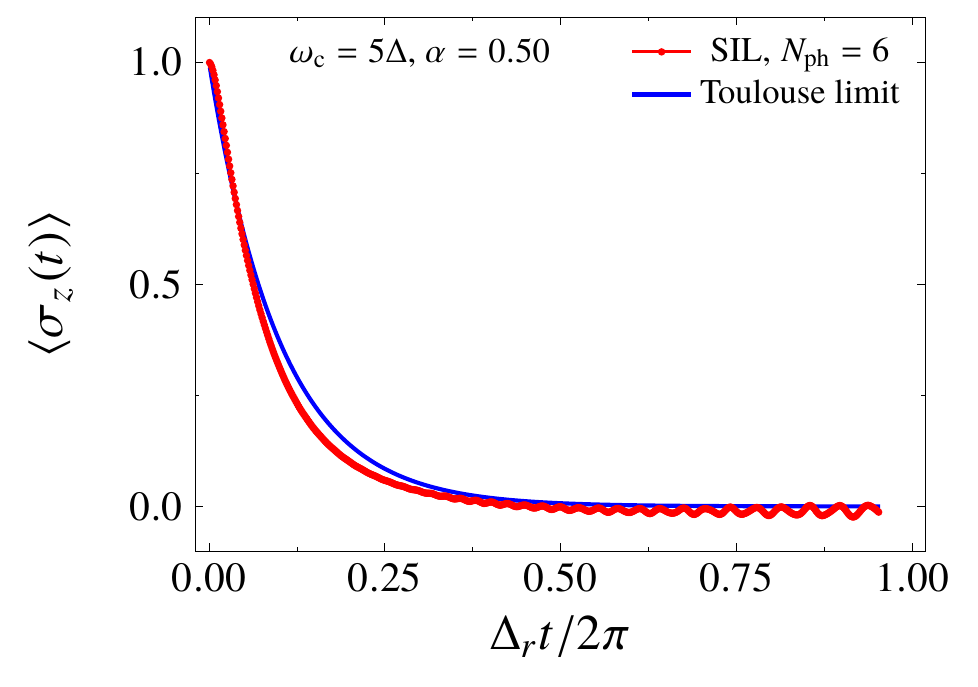}
		\caption{Magnetization $\ev{\sigma_z(t)}$ as a function of the rescaled time, computed at the Toulouse point ($\alpha=1/2$), for $h_0=0$ (unbiased case), at $T=0$; we plot of the numerical SIL result (red curve), compared with the theoretical curve in Eq.~\eqref{eq:Toulouse} (solid blue curve), which is valid in the limit $\omega\ped{c}\to\infty$. As in the main text, $M=50$, $\omega\ped{c}=5\Delta$, $N\ped{ph}=6$.}
		\label{fig:toulcomparison}
	\end{figure}

	In the following, we compare our numerical findings for the SBM with well known theoretical results from the literature. As discussed in the main text, in the unbiased case approximate analytical treatments (NIBA) have been devised in order to describe the underdamped oscillation in time of the qubit magnetization and its crossover to the incoherent regime. Although it is argued that these theories generally don't give the correct analytical expression for the oscillation frequency, the result for the quality factor of the oscillation, in the limit of large $\omega\ped{c}$ and $T=0$, reads               
    \begin{equation}\label{eq:qualityfactor}
    Q=\frac{\Omega}{\gamma}= \cot(\frac{\pi \alpha}{2(1-\alpha)})    
    \end{equation} 	
    In Fig.~\ref{fig:qf} we plot the quality factor computed by fitting the numerical curves of Sec.~\ref{sec:Toulouse} against the theoretical result in Eq.~\eqref{eq:qualityfactor}, as a function of the coupling strength $\alpha$ in the range $\numrange[range-phrase = \text{~to~},exponent-product=\cdot]{1.0e-1}{4.5e-1}$. The numerical results fairly agree with the theoretical prediction, showing that our technique can successfully describe the dissipative behavior expected from conventional theories.
     However, a direct comparison with  Eq.~\eqref{eq:Toulouse} for $h_0=0$ shows that at the Toulouse point our numerical simulations cannot correctly describe the expected result in the whole time domain. As it can be derived from Fig.~\ref{fig:toulcomparison}, a deviation from the analytical result can be observed in the region of intermediate times, while 
	at longer times a residual coherent behavior can be observed which cannot be found in non-perturbative analytical treatments. 
	
\begin{figure}[tb]
		\centering
		\includegraphics[width=\linewidth]{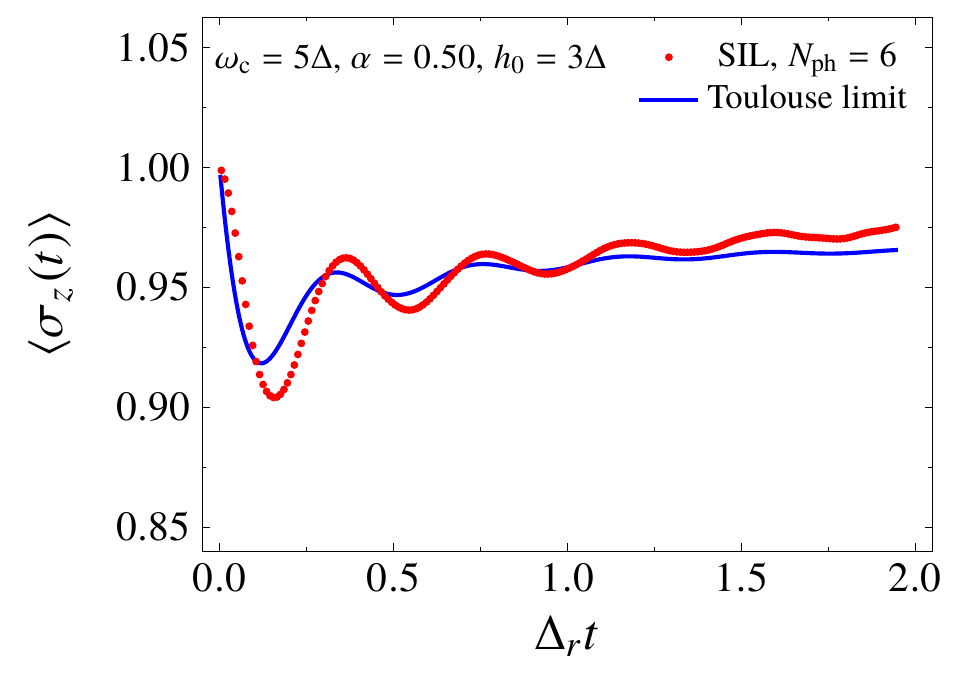}
		\caption{Magnetization $\ev{\sigma_z(t)}$ as a function of the rescaled time, computed at the Toulouse point ($\alpha=1/2$), with fixed bias $h_0=3\Delta$, and $T=0$; we plot the numerical SIL result (red points), compared with the theoretical curve in Eq.~\eqref{eq:Toulouse} (solid blue curve). As in the main text, $M=50$, $\omega\ped{c}=5\Delta$,$N\ped{ph}=6$.}
		\label{fig:toulbiased}
	\end{figure}

	Similar results can be found in the biased case: as it is evident from Fig.~\ref{fig:toulbiased}, while the numerical curve correctly describe the qualitative behavior of the function in Eq.~\eqref{eq:Toulouse}, which is strictly valid for $\omega\ped{c}\rightarrow \infty$ as leading order, several quantitative differences can be observed,~\eg~the long-time value of $\ev{\sigma_z(t)}$ slightly differs from that expected from Eq.~\eqref{eq:Toulouse} at $T=0$,~\ie~$\ev{\sigma_z(\infty)}=\frac{2}{\pi}\arctan(\frac{4h_0\omega_c}{\pi\Delta^2})$.
	As anticipated in the main text, these results are mainly due to the small value of the frequency cutoff $\omega\ped{c}$ chosen: we expect these small numerical differences to vanish as long as the frequency cutoff, as well as the number of the bath oscillators $M$ is increased.  
	However, as the dimension of the truncated Hilbert space considered rapidly grows with the absolute maximum number of excitations $N\ped{ph}$ and the number of modes $M$ (see App.~\ref{app:SIL} for details), the inclusion of additional modes in the strong coupling regime can become prohibitively costly. These findings point towards the need for an optimized basis of states for the implementation of SIL method, which could hopefully reduce its computational cost.             
     	
	\section{SIL method}\label{app:SIL}
	    
     The dissipative dynamics governed by Eq.~\eqref{eq:TotalHam} can be studied numerically by evaluating the evolution operator $U(t, t_{0})$ of the whole qubit+bath system, in order to simulate the unitary dynamics of the global state $\ket{\Psi(t)}$ of the system and then carry out the trace over the bath degrees of freedom. This task can be performed by employing a discretization of the bath modes entering in Eq.~\eqref{eq:BathHam}, a suitable truncation scheme of the bath Hilbert space, followed by the application of SIL method \cite{Cangemi:SIL,Lanczos1,Lanczos2}.  
     The discretization of the bath modes can be performed by choosing a density of states $\rho(\omega)$ and by fixing the total number of bosonic modes $M$ in the range $[0,2\omega\ped{c}]$; here we adopt an exponentially decreasing density of states with frequency cutoff $\omega\ped{c}$
     \begin{equation}
     \rho(\omega)\propto\exp(-\frac{\omega}{\omega\ped{c}})\mbox{ , } \int\ped{0}^{\omega\ped{c}}\rho(\omega)=M
     \end{equation}    
     As a consequence, for each mode of frequency $\omega_k$ we choose the coupling strength $g(\omega_k)$ obeying to 
     \begin{equation}
     \rho(\omega_k)g^2(\omega_k)=2 \alpha \frac{\omega_k^{s}}{\omega\ped{c}^{s-1}}\eu^{-\frac{\omega_k}{\omega\ped{c}}}
     \end{equation}    
     It is clear that this finite system can mimick the theoretical model of a continuum set of modes constituting tha bath as long as $M$ is sufficiently high. Every bath state is described by a set of basis states $ \Set{\ket{n_{1},n_{2},\dots,n_{M}}} $, where  $ n_{k} $ is the occupation number of the $k$-th bosonic mode of the bath. In order to perform a truncation of the space spanned by these states, we fix the absolute maximum number of bosonic excitations $N\ped{ph}$ with respect to the thermal equilibrium, and we restrict the description only to states for which $ \Delta n_{k}=n_{k}-n_{k}\api{eq} = \Set{0, \pm 1, \pm 2,\dots, \pm N\ped{ph}}$, with $\sum_k \abs{\Delta n_{k}}\leq N\ped{ph}$, where $n_{k}\api{eq}$ is the occupation number of the $k$-th bosonic mode at equilibrium. Hence this numerical approach can give an exact description of the physics up to terms in $\alpha^{N\ped{ph}}$. The resulting dimension of the truncated Hilbert space of the qubit+bath system is thus equal to    
     \begin{equation}
     \mathcal{N}=2\sum_{j=1}^{N\ped{ph}}\binom{N\ped{ph}}{j}\binom{M}{j}
     \end{equation} 
     After having fixed the set of basis states, we compute iteratively the state of the system $\ket{\Psi(t)}$ at each time $t$: it can be achieved by employing a discretization of the total evolution time interval in steps $\dd{t}$, and a projection of the Hamiltonian evaluated at midpoint in each time interval $\rng{t}{t+\dd{t}}$ into the $ n $-dimensional subspace $\mathcal{K}=\Set{\ket{\Psi(t)}, \ham\ket{\Psi(t)}, \dots, \ham^{n}\ket{\Psi(t)}}$ spanned by the Krylov orthonormal vectors $\Set{\ket{\Phi_{k}}}_{k=1}^{n}$, which we coumpute using recursive Gram-Schmidt orthogonalization techniques. The reduced Hamiltonian reads $\tilde{\ham}(t+\dd{t}/2)=P \ham(t+\dd{t}/2) P^{\dagger}$, where $P$ is the projection operator in the subspace $\mathcal{K}$; it can be easily diagonalized and the evolution operator in terms of the eigenstates of $\tilde{\ham}(t+\dd{t}/2)$ can be derived       
     \beq\label{eq:projectedEv}
     \tilde{U}(t+\dd{t},t)\simeq \exp[-\iu \tilde{\ham}(t+\dd{t}/2) \dd{t}].
     \eneq   
     Finally, we expand the state at previous time $t$ $\ket{\Psi(t)}$ in terms of the eigenvectors of $\tilde{\ham}\qty(t+\dd{t}/2)$, and thus we are able to compute the state at the end of the time interval $\ket{\Psi(t+dt)}$ using \eqref{eq:projectedEv} by means of matrix products. The computation of the full ket state allows us to derive the density matrix of the  system, from which we can numerically trace over the bath degrees of freedom and compute the reduced density matrix of the qubit. Every bath observable can also be computed.

	\section{Qubit dynamics in the absence of dissipation}\label{app:unitary}
		
		The qubit dynamics ruled by Eq.~\eqref{eq:qubitperiodic} can be easily solved in the counter-rotating reference frame around the $\hat{y}$ axis, if the static field $h\ped{0}$ is taken to be equal to zero. Given the rotation operator of angle $\phi$ around the $\hat{n}$ direction $U(\hat{n},\phi)=\exp(-\iu\boldsymbol{\hat{n}}\cdot\boldsymbol{\sigma}\phi/2)$, we can write the Schr\"odinger equation for the rotated ket $\ket{\psi(t)}_r=U\ket{\psi(t)}$; taking $\hbar=1$, the Hamiltonian $H_r$ in the rotating frame reads
		\begin{equation}
        H_r=\iu\derivative{U}{t}U^\dagger + U H U^\dagger 		
		\end{equation}    
		Notice that the Hamiltonian can be written as a sum of two terms, the first is the adiabatic gauge potential in the rotating frame, while the second is the diagonalized Hamiltonian operator. The adiabatic gauge term is responsible for the transitions between diabatic states in the rotating frame: this implies that, as shown in \cite{Berry,SelsE3909}, at least in principle it is possible to engineer counter-adiabatic Hamiltonians for which these transitions are always suppressed in the rotating frame.   
		In our conventional scheme, we take $\hat{n}=\hat{y}$ and impose the counter-rotating condition $\phi(t)=-\theta(t)$, the resulting Hamiltonian $H_r$ is time-independent and it reads:
		\begin{equation}\label{AdiabaticHam}
		H\ped{r}=- \frac{h}{2} \sigma_z -\frac{\dot{\theta}}{2}\sigma_y 
		\end{equation}
		The adiabatic eigenvalues of Hamiltonian in Eq.~\eqref{AdiabaticHam} are $E_\pm=\pm\frac{1}{2}\sqrt{\dot{\theta}^2 + h^2}$; after computing the adiabatic eigenvectors of Eq.~\eqref{AdiabaticHam}, given the initial state of the qubit $\ket{\psi(t\ped{0})}$, the state of the qubit at final time $t$ can be easily found:
		\begin{equation}\
		\ket{\psi(t)}=U^\dagger(\hat{y},\theta(t)) U_r(t,t\ped{0})U(\hat{y},\theta(t_{0}))\ket{\psi(t_0)}
		\end{equation}    
		where $U\ped{r}(t,t\ped{0})$ is the evolution operator in the rotating frame. In the protocol described in \ref{sec:Magnetic} the qubit is initially prepared in the state $\ket{\psi(t\ped{0})}=\ket{\hat{z}; +}$: by choosing $t_{0}=0$, the magnetic field evolves from $\theta(0)=0$ to $\theta(t\ped{f})=\pi$, thus the final state reads    
		\begin{equation}\label{eq:finalket}
		\ket{\psi(t\ped{f})}=-\iu\sigma_y\exp(-\iu H\ped{r} t\ped{f})\ket{+}			
		\end{equation} 
		From Eq.~\eqref{eq:finalket}, the magnetization along $\hat{z}$ at the end of the protocol can be straightforwardly derived, and it reads:
		\begin{equation}
        \ev{\sigma_z(t\ped{f})}= - \frac{h^2 + \dot{\theta}^2 \cos(\pi\sqrt{\dot{\theta}^2 + h^2}/\dot{\theta})}{\dot{\theta}^2 + h^2}
        \end{equation} 
        The excess energy at the end of the annealing can thus be directly evaluated and gives Eq.~\eqref{eq:eresclosed}.  
     
\bibliography{paper}   
         
\end{document}